\documentclass[aps,prl,twocolumn,groupedaddress, showpacs]{revtex4}

\usepackage{graphicx,amsmath,amssymb,bbm}
\usepackage{dcolumn}
\usepackage{bm}
\usepackage[sort&compress]{natbib}

\newcommand{\rd}{{\rm d}}
\begin{document}

\title{Accurate determination of elastic parameters for multi-component membranes}

\author{Stefan Semrau(1*)}
\author{Timon Idema(2*)}
\author{Laurent Holtzer(1)}
\author{Thomas Schmidt(1)}
\author{Cornelis Storm(2)}
\affiliation{
1 Physics of Life Processes; 2 Instituut-Lorentz for Theoretical Physics\\
Leiden Institute of Physics, Leiden University, P.O.Box 9506, 2300 RA Leiden, The Netherlands\\
(*) authors contributed equally
}

\date{\today}

\begin{abstract}
Heterogeneities in the cell membrane due to coexisting lipid phases
have been conjectured to play a major functional role in cell
signaling and membrane trafficking. Thereby the material properties
of multiphase systems, such as the line tension and the bending
moduli, are crucially involved in the kinetics and the asymptotic
behavior of phase separation. In this Letter we present a combined
analytical and experimental approach to determine the properties of
phase-separated vesicle systems. First we develop an analytical
model for the vesicle shape of weakly budded biphasic vesicles.
Subsequently experimental data on vesicle shape and membrane
fluctuations are taken and compared to the model. The combined
approach allows for a reproducible and reliable determination of the
physical parameters of complex vesicle systems. The parameters obtained
set limits for the size and stability of nanodomains in the plasma
membrane of living cells.
\end{abstract}

\pacs{87.15.Ya, 87.16.Dg, 87.17.Aa, 02.40.Hw}

\maketitle
The recent interest in coexisting phases in lipid
bilayers originates in the supposed existence of lipid
heterogeneities in the plasma membrane of cells. A significant role
in cell signaling and traffic is attributed to small lipid domains
called ``rafts''~\cite{rafts, lommerse05}. While their existence
in living cells remains the subject of lively
debate, micrometer-sized domains are readily reconstituted in
giant unilamellar vesicles (GUVs) made from binary or ternary lipid
mixtures~\cite{dietrich01}. Extensive studies of these and similar
model systems have brought to light a rich variety of phases, phase
transitions and coexistence regimes~\cite{veatch05}. 
In contrast to these model systems, no large (micrometer sized) membrane domains have been
observed in vivo. If indeed phase separation occurs in vivo,
additional processes which can arrest it prematurely must be considered.
It has been suggested that nanodomains might be stabilized by
entropy~\cite{frolov06} or that, alternatively, active cellular
processes are necessary to control the domain size~\cite{turner05}. A
third explanation is that curvature-mediated interactions might
conspire to create an effective repulsion between domains, impeding
and ultimately halting their fusion as the phase separation
progresses. Each of these three processes depends critically on
membrane parameters such as line tension~\cite{lipowski92}, curvature moduli and even
the elusive Gaussian rigidities~\cite{julicher93}. Although some studies report values~\cite{baumgart05} or upper bounds~\cite{siegel04, allain04} for
these membrane parameters, a systematic method to determine them from experiments that does not require extensive numerical simulation and fitting is lacking.
\begin{figure}[ht]
\includegraphics*[width=.8\columnwidth]{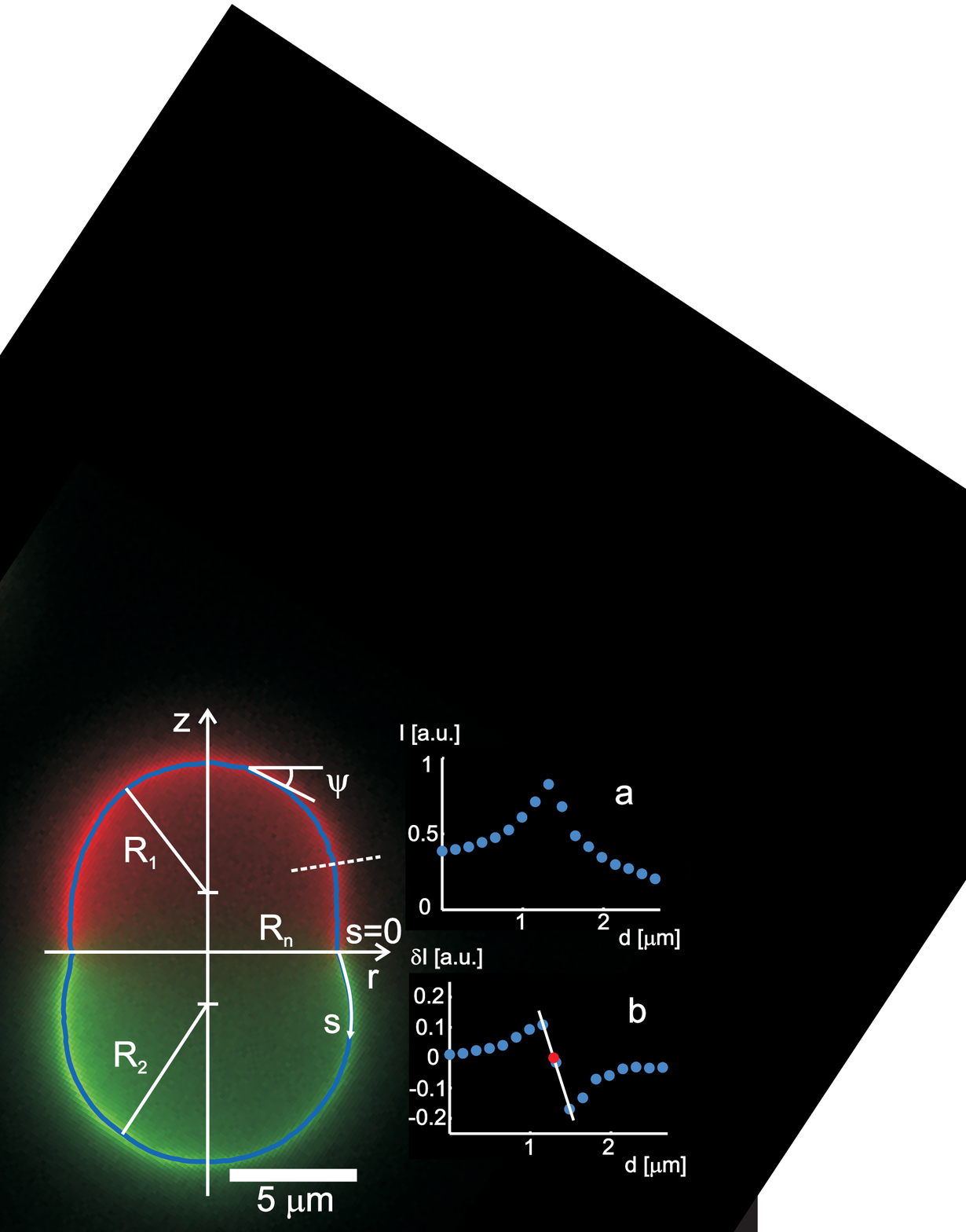}
\vspace{-0.3cm}
\caption{\label{fig:vesicle}
Fluorescence raw data (red: $L_{\mbox{\scriptsize o}}$ domain, green: $L_{\mbox{\scriptsize d}}$ domain) with superimposed contour (light blue). Insets: principle of contour fitting; 
a: intensity profile normal to the vesicle contour (taken along the dashed line in the main image); b: first derivative of the profile with linear fit around the vesicle edge (white line). The red point marks the vesicle edge.
\vspace{-0.7cm}
}
\end{figure}
We present here a straightforward fully analytical method that allows for a precise, simultaneous determination of the line tension, the bending rigidity and
the difference in Gaussian moduli from biphasic GUVs. Both the liquid ordered~$L_{\mbox{\scriptsize o}}$ and the liquid disordered~$L_{\mbox{\scriptsize d}}$ phase are quantitatively 
characterized with high accuracy. Our method relies on an analytical expression for the shape of a moderately budded vesicle.
A one-parameter fit to experimental shapes permits unambiguous determination of the line tension and the difference in Gaussian moduli. 
Our results provide important clues as to the origin and magnitude of long-ranged membrane-mediated interactions, which have been proposed 
recently as an explanation for the trapped coarsening~\cite{yanagisawa07, reynwar07} and the very regular domain structure of a meta-stable state~\cite{baumgart03} 
found in experiments. Furthermore, our results show that nanometer-sized phase separated domains will be stable in
life cells.

\paragraph{\label{sec:model}Model}
The free energy associated with the bending of a thin membrane is described by the Canham-Helfrich free energy~\cite{helfrich73}. We ignore any spontaneous curvature of the membrane because the experimental system has ample time to relax any asymmetries between the leaflets. For a two-component vesicle with line tension~$\tau$ between the components, the free energy then reads:
\vspace{-0.2cm}
\begin{equation}
\label{Helfrich}
\mathcal{E} = \sum_{i=1,2} \int_{S_i} \left( 2
\kappa_i H^2 + \kappa_{\mbox{\tiny G}}^{(i)} K + \sigma_i \right) \rd A - pV + \tau
\oint_{\partial S} \rd\ell,
\end{equation}
where the~$\kappa_i$ and~$\kappa_{\mbox{\tiny G}}^{(i)}$ are the bending and Gaussian moduli of the two phases, respectively, the~$\sigma_i$ are their surface tensions, 
and $p$ is the internal pressure. In equilibrated shapes such as our experimental vesicles, the force of the internal Laplace pressure is compensated by the surface tensions; consequently, both contributions drop out of the shape equations~\cite{variationalcalculus}.
For each phase, we integrate the mean~($H$) and Gaussian~($K$) curvature over the membrane patch~$S_i$ occupied by that phase; the line tension contributes at the boundary~$\partial S$ of the two phases. Using the Gauss-Bonnet theorem, we find that the Gaussian curvature term yields a constant bulk contribution plus a boundary term~\cite{carmo76}.

The axisymmetric shapes of interest (Fig.~\ref{fig:vesicle}) are fully described by the contact angle~$\psi$ as a function of the arc length~$s$
along the surface contour. The coordinates~$(r(s),z(s))$ are fixed
by the geometrical conditions $\dot r = \cos \psi(s)$ and $\dot z =
-\sin \psi(s)$, where dots denote derivatives with respect to the
arclength. Variational calculus gives the basic shape
equation~\cite{variationalcalculus}:
\begin{equation}
\label{shapeeq} \ddot \psi \cos \psi = -\frac12 \dot\psi^2 \sin\psi
-\frac{\cos^2\psi}{r}\dot\psi + \frac{\cos^2\psi+1}{2 r^2}\sin\psi.
\end{equation}
This equation holds for each of the phases separately. The radial coordinate~$r(s)$ and tangent angle~$\psi(s)$ must of course be continuous at the domain boundary. Additionally, the variational derivation of equation~(\ref{shapeeq}) gives two more boundary conditions~\cite{julicher96}:
\vspace{-0.2cm}
\begin{eqnarray}
\label{psidotbc}
\lim_{\varepsilon \downarrow 0} (\kappa_2 \dot\psi(\varepsilon) - \kappa_1 \dot\psi(-\varepsilon)) &=& (\Delta\kappa+\Delta\kappa_{\mbox{\tiny G}}) \frac{\sin\psi_0}{R_n},\\
\label{psiddotbc}
\lim_{\varepsilon \downarrow 0} (\kappa_2 \ddot\psi(\varepsilon) - \kappa_1 \ddot\psi(-\varepsilon)) &&\nonumber\\
&&\hspace{-95pt} = -(2\Delta\kappa+\Delta\kappa_{\mbox{\tiny G}})\frac{\cos\psi_0 \sin\psi_0}{R_n^2} + \frac{\sin\psi_0}{R_n}\tau,
\vspace{-0.2cm}
\end{eqnarray}
with~$R_n$ and~$\psi_0$ the vesicle radius and tangent angle at the domain boundary, $\Delta \kappa = \kappa_1-\kappa_2$, $\Delta \kappa_{\mbox{\tiny G}} = \kappa_{\mbox{\tiny G}}^{(1)}-\kappa_{\mbox{\tiny G}}^{(2)}$, 
and the domain boundary located at~$s=0$.

The sphere is a solution of the shape equation~(\ref{shapeeq}); we can therefore use it as an ansatz for the vesicle shape far from the domain boundary. We split the
vesicle into three parts: a neck domain around the domain boundary, where
the boundary terms dominate the shape, and two bulk domains, where
the solution asymptotically approaches the sphere.
Perturbation analysis, performed by expanding Eq.~(\ref{shapeeq}) around the
spherical shape, then gives for the bulk domains:
\vspace{-0.2cm}
\begin{equation}
\label{psibulk} \psi_\mathrm{bulk}^{(i)}(s) =
\frac{s+s_0^{(i)}}{R_i} + c_i R_i
\log\left(\frac{s}{s_0^{(i)}}\right).
\end{equation}

Here~$R_i$ is the radius of curvature of the underlying sphere and $s_0^{(i)}$ the distance (set by the area constraint on the vesicle) from the point $r=0$ to the domain boundary. 
As was shown by Lipowski~\cite{lipowski92}, the governing length scale is the {\em invagination length}, defined as the ratio~$\xi_i=\kappa_i/\tau$ of the bending modulus and the line tension. Our three-domain approach applies when this invagination length is small compared to the size of the vesicle. At $s=\xi_i$ the line tension, rather than the bending modulus, becomes the dominant term in the energy. Self-consistency of the solution requires that the deviation from the sphere solution at that point be small, i.e. given by the dimensionless quantity~$\xi_i/R_i$. This fixes the integration constant~$c_i$.

Near the domain boundary, $\psi$ must have a local extremum in each of the phases and we can therefore expand it as
\vspace{-0.2cm}
\begin{equation}
\label{psineck} \psi_\mathrm{neck}^{(i)}(s) = \psi_0^{(i)} + \dot
\psi_0^{(i)} s + \frac12 \ddot \psi_0^{(i)} s^2.
\end{equation}

These neck solutions must match at the domain boundary and also satisfy conditions~(\ref{psidotbc},~\ref{psiddotbc}). Furthermore, they also need to match 
the bulk solutions to ensure continuity of $\psi$ and its derivative $\dot\psi$. In total this yields seven equations for the eight unknowns~$\{\psi_0^{(i)}, \dot\psi_0^{(i)},
\ddot\psi_0^{(i)}, s_i\}$. The necessary eighth equation is provided by the condition of
continuity of~$r(s)$ at the domain boundary.

\begin{figure}[ht]
\includegraphics[width=.8\columnwidth]{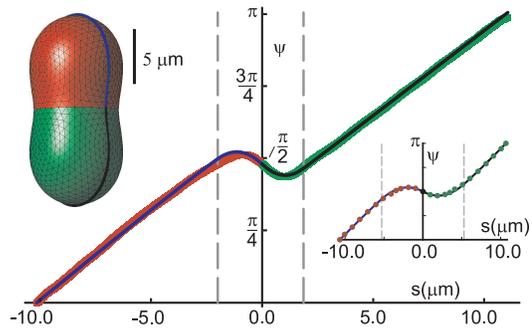}
\vspace{-0.3cm}
\caption{\label{fig:datafit}  Example for an experimentally obtained
$\psi(s)$ plot (red: $L_{\mbox{\scriptsize o}}$ phase, green: $L_{\mbox{\scriptsize d}}$ phase ) together
with the best fit of the model (blue: $L_{\mbox{\scriptsize o}}$ phase, black: $L_{\mbox{\scriptsize d}}$
phase). The dashed lines mark the transition points between the neck and bulk domains. Insets: Fit to a numerically obtained shape (using Surface Evolver).}
\vspace{-0.7cm}
\end{figure}

Put together the neck and bulk components of~$\psi$ give a vesicle solution for specified values of the material parameters~$\{\kappa_i, \Delta \kappa_{\mbox{\tiny G}}, \tau\}$. 
This solution compares extremely well to numerically determined shapes (obtained using the Surface Evolver package~\cite{brakke92}, Fig.~\ref{fig:datafit}).
Moreover, for the symmetric case of domains with identical values of $\kappa$, we can compare to earlier modeling in Ref.~\cite{julicher93}. The vesicle can then be described by a single dimensionless parameter $\lambda = \xi/R_0$, where $4\pi R_0^2$ equals the vesicle area. The `budding transition' (where the broad neck destabilizes in favor of a small neck) is numerically found in Ref.~\cite{julicher93} to occur at $\lambda=4.5$ for equally sized domains; our model gives a value of $\lambda=4.63$.

\paragraph{Experiment}
Giant unilamellar vesicles (GUVs) were produced by electroformation
from a mixture of 30~\% DOPC, 50~\% brain sphingomyelin, and 20~\% cholesterol at 55~$^\circ$C. Subsequently lowering the temperature to 20~$^\circ$C resulted 
in the spontaneous formation of liquid ordered~$L_{\mbox{\scriptsize o}}$ and liquid disordered~$L_{\mbox{\scriptsize d}}$ domains on the vesicles.
The $L_{\mbox{\scriptsize d}}$ phase was stained by a small amount of Rhodamine-DOPE (0.2~\%). In order to stain the $L_{\mbox{\scriptsize o}}$ phase a small amount (0.2~\%) of the 
ganglioside GM1 was added, and subsequently choleratoxin labeled with Alexa~647 was bound to the GM1~\cite{APLstuff}. For imaging we chose a wide-field 
epi-fluorescence setup~\cite{lommerse05} because short illumination times (1-5 ms) prevent shape
fluctuations with short correlation times from being washed out.
The raw data of a typical vesicle is shown in Fig.~\ref{fig:vesicle}.
The lateral resolution of the equatorial optical sections was limited by diffraction and pixelation effects. In the normal
direction, however, a high (sub-pixel) accuracy was obtained. The
upper inset in Fig.~\ref{fig:vesicle} shows a typical intensity
profile along a line perpendicular to the contour. We determine
numerically the profile's first derivative (lower inset in
Fig.~\ref{fig:vesicle}) and fit the central part around the maximum
intensity with a straight line. The intercept with the $x$-axis gives
the position of the vesicle edge. The positional accuracy achieved
is typically 20 nm. The contours obtained were subsequently
smoothed by a polynomial and all contours from the same vesicle
(typically around 1000) were averaged to give the final result for
the mean contour.\vspace{-0.3cm}

\begin{figure}[ht]
\includegraphics[width=.8\columnwidth]{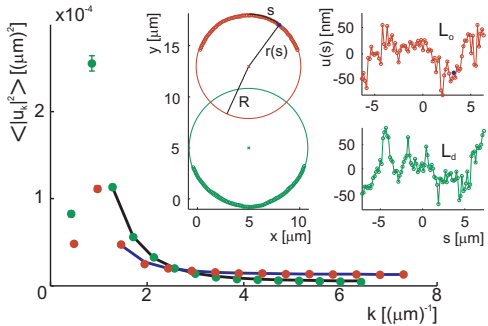}
\vspace{-0.3cm}
\caption{ \label{fig:flucspec} Fluctuation spectra of the ordered
(red circles) and disordered (green circles) domains. The
corresponding best fits of Eq.~\eqref{eq:flucspec} are shown in blue
and black respectively. Inset: Typical real-space fluctuations along
the vesicle perimeter.}
\end{figure}
\vspace{-0.3cm}
Spectra of the shape fluctuations were obtained from those
parts of the contours that were nearly circular, i.e. far away from
the neck domain.
Fluctuations were determined for each single contour as the
difference between the local radius $r$ and the ensemble averaged
radius $R$ of a circle fitted to patches around the vesicles' poles:
$u(s) = r(s) - R$ where $s$ is the arclength along the circle, see
Fig.~\ref{fig:flucspec}.
The experimental fluctuation spectrum was obtained by Fourier transform as $u_{k} = \frac{1}{a} \int_{-a/2}^{a/2}\rd s \;  r(s) e^{-ik \cdot s}$, where $a$ is the
arclength of the contour patch, and $k = n \cdot \frac{2 \pi}{a}$ with
$n$ a non-zero integer.
Taking into account the finite patch size~\cite{mutz90} and following the
spectral analysis of a closed vesicle shell developed by
P\'ecr\'eaux et al.~\cite{pecreaux04} leads to a power spectrum for the
vesicle fluctuation
\vspace{-0.2cm}
\begin{equation}
\overline{\langle | u_{k}|^2 \rangle} =   \sum_{q} \left( \frac{\sin
((k-q)\frac{a}{2})}{(k-q)\frac{a}{2}} \right)^2 \overline{\langle |
u_{q}|^2 \rangle}_{\mbox{\scriptsize sph}}.
\label{eq:flucspec}
\end{equation}
Here $q = \frac{2 \pi}{L} m $ with $m$ a non-zero integer, $L = 2
\pi R$, and $\overline{\langle | u_{q}|^2 \rangle}_{\mbox{\scriptsize
sph}}$ the spectrum of the entire vesicle derived in~\cite{pecreaux04}, where the overline indicates temporal averaging during the illumination time.
Eq.~(\ref{eq:flucspec}) was derived from the Canham-Helfrich
free energy for a flat membrane with periodicity $L$. However, as
shown in \cite{pecreaux04}, the spectrum of a sphere with radius~$R$
differs from that of the flat membrane only for the lowest wave
numbers~$k$. Therefore we can use Eq.~(\ref{eq:flucspec}) to fit our
fluctuation spectra if we omit the two lowest modes. Examples of such
fits are shown in Fig.~\ref{fig:flucspec}.

\paragraph{Results}
Fits of the fluctuation spectra using Eq.~(\ref{eq:flucspec})
give the values of the bending moduli and surface tensions of the
two phases. Using these values, we fit the experimentally
obtained vesicle shapes with the model described above. This leaves
us with two parameters: the line tension~$\tau$ between the two
phases and the difference~$\Delta \kappa_{\mbox{\tiny G}}$ between their Gaussian moduli. 
Since the experimental data show that $\psi$ at
the domain boundary follows a straight, continuous line we further
assume that the derivative $\dot \psi$ is continuous at the domain boundary (as suggested before~\cite{julicher93, julicher96}).
Imposing this additional condition fixes the value of~$\Delta \kappa_{\mbox{\tiny G}}$ for given~$\tau$, leaving us with a single free parameter to fully describe 
the system~\cite{twoparameterfit}. By fitting the experimental data, we directly
extract the line tension. An example fit is shown in Fig.~\ref{fig:datafit}. Values found for the
bending moduli are~$8 \pm 1 \cdot 10^{-19}$ J for the $L_{\mbox{\scriptsize o}}$ domain
and $1.9 \pm 0.5 \cdot 10^{-19}$ J for the $L_{\mbox{\scriptsize d}}$ domain. For the line
tension we found a value of~$1.2 \pm 0.3$ pN, which is in the same
range as that estimated by Baumgart et al.~\cite{baumgart03}.
Finally, the difference in Gaussian moduli is about~$3 \pm 1 \cdot
10^{-19}$ J, in accordance with the earlier established upper bound ($\kappa_{\mbox{\tiny G}} \leq -0.83 \kappa$) reported by Siegel and Kozlov~\cite{siegel04}. 
An overview of the results is given in Table~\ref{table:results}.

\paragraph{Discussion}
Ultimately, one worries about the membrane's elastic parameters because their precise magnitude has important consequences for the morphology 
and dynamics of cells. The literature is replete with theoretical speculations which depend strongly on, among others, the line tension. 
While the values we report apply to reconstituted vesicles, we can nonetheless use them in some of these models to explore possible implications for cellular membranes.
The majority of the investigated vesicles finally evolved into the
fully phase separated state. This finding is in agreement with
previous work by Frolov et al.~\cite{frolov06}, which predicts, for
line tensions larger than 0.4 pN, complete phase separation for
systems in equilibrium.
It should be noted that the line tension
found is also smaller than the critical line tension leading to
budding: recent results by Liu et al.~\cite{liu06} show that for
endocytosis by means of membrane budding both high line tensions
($>10$ pN) and large domains are necessary. Therefore nanodomains
will be stable and will not bud.
In cells, however, additional mechanisms must be considered. To explain the absence of large domains in vivo, Turner et al.~\cite{turner05} 
make use of a continuous membrane recycling mechanism. For the membrane parameters we have determined such a mechanism predicts asymptotic 
domains of $\sim$10 nm in diameter. Our results, in combination with
active membrane recycling therefore support a minimal physical
mechanism as a stabilizer for nanodomains in cells.
A separate effect, purely based on the elastic properties of
membranes may further stabilize smaller domains in vivo. Recently,
Yanagisawa et al. explored the consequences of a repulsive
interaction between nearby buds~\cite{yanagisawa07} and reported
that such interactions can arrest the phase separation kinetics. The
elastic perturbations induced by domains in the membrane, as described in this Letter, are obvious candidates for producing additional 
interactions between buds at any distance, further assisting in the creation of such a kinetic arrest.
As M\"uller et al. have shown for a flat membrane, two distortions on
the same side of an infinite flat membrane repel on all length
scales~\cite{muller05}. The experimental observation of multiple domains ordered in (quasi-)crystalline fashion in model membranes~\cite{baumgart03} 
strongly suggests a similar repulsive interaction in spherical vesicle systems. This is indeed evidenced by preliminary numerical exploration of this 
system using Surface Evolver~\cite{brakke92}.
It is of course straightforward to adapt the scheme outlined above
to include long-range interactions between transmembrane proteins
that impose a curvature on the membrane, e.g. scaffolding
proteins~\cite{reynwar07, scaffoldingproteins}. Membrane mediated
interactions act over length scales much larger than Van der Waals
or electrostatic interactions and could provide an alternative or
additional physical mechanism for processes like protein clustering
and domain formation. Our results and methods allow not only to
determine the parameters relevant to processes like these, but also
give a practical analytical handle on the shapes involved. This, in
turn, will help decide between competing proposals for mechanisms
involving membrane bending: protein interactions, endocytosis and
the formation and stabilization of functional membrane domains.

\paragraph{Acknowledgments}
This work was supported by funds from the Netherlands Organization for Scientific Research (NWO-FOM) within the program on Material Properties of Biological Assemblies (FOM-L1707M \& FOM-L2601M).
\vspace{-0.7cm}

\begin{table}[t]
{\footnotesize
\begin{tabular}[t]{|c|c|c|c|c|c|c|}
\hline
 & $\sigma_{\mbox{\scriptsize d}}$ & $\kappa_{\mbox{\scriptsize d}}$ & $\sigma_{\mbox{\scriptsize o}}$ & $\kappa_{\mbox{\scriptsize o}}$ & $\tau$ & $\Delta \kappa_{\mbox{\tiny G}}$\\
 &($10^{-7} \frac{N}{m}$) & ($10^{-19}$J) & ($10^{-7} \frac{N}{m}$) & ($10^{-19}$J) & (pN) & ($10^{-19}$J)\\[0.1 cm]
\hline
 1 & 2.8 $\pm$ 0.2 & 2.2 $\pm$ 0.1 & 0.3 $\pm$ 0.3 & 8.0 $\pm$ 1.3 & 1.5 $\pm$ 0.3 & 2.5 $\pm$ 2 \\

 2 & 5.8 $\pm$ 0.5 & 1.8 $\pm$ 0.2 & 2.1 $\pm$ 0.4 & 8.2 $\pm$ 1.5 & 1.2 $\pm$ 0.4 & 2.0 $\pm$ 2 \\

 3 & 3.5 $\pm$ 0.3 & 2.0 $\pm$ 0.1 & 2.0 $\pm$ 0.5 & 8.2 $\pm$ 1.4 & 1.2 $\pm$ 0.3 & 2.5 $\pm$ 2 \\

 4 & 2.8 $\pm$ 0.2 & 1.9 $\pm$ 0.1 & 2.5 $\pm$ 0.5 & 8.3 $\pm$ 1.2 & 1.2 $\pm$ 0.4 & 4.0 $\pm$ 2 \\

 5 & 2.3 $\pm$ 0.1 & 1.6 $\pm$ 0.1 & 0.6 $\pm$ 0.3 & 8.0 $\pm$ 1.6 & 1.1 $\pm$ 0.5 & 4.0 $\pm$ 3 \\
\hline
\end{tabular}
}
\caption{\label{table:results} Values of the material parameters for five different vesicles. The surface tensions and bending moduli of the $L_{\mbox{\scriptsize d}}$ and $L_{\mbox{\scriptsize o}}$ phase are determined from the fluctuation spectrum; the line tension and difference in Gaussian moduli are subsequently determined 
using our analytical model.
\vspace{-.5cm}
}
\end{table}

\bibliographystyle{unsort}

\begin{thebibliography}{25}
\vspace{-0.7cm}
\bibitem{rafts}
K. Simons and E. Ikonen, Nature {\bf 387}, 569 (1997); R. G. W. Anderson and K. Jacobson, Science {\bf 296}, 1821 (2002); M. Edidin, Ann. Rev. Biophys. Biomol. Struct. {\bf 32}, 257 (2003); F. R. Maxfield and I. Tabas, Nature {\bf 438}, 612 (2005); J. F. Hancock, Nature Rev. Mol. Cell Biol. {\bf 7}, 456 (2006).

\bibitem{lommerse05}
P. H. M. Lommerse, B. E. Snaar-Jagalska, H. P. Spaink, and T. Schmidt, J. Cell. Sci. {\bf 118}, 1799 (2005).

\bibitem{dietrich01}
C. Dietrich \emph{et al.}, Biophys. J. {\bf 80}, 1417 (2001).

\bibitem{veatch05}
S. L. Veatch and S. L. Keller, Phys. Rev. Lett. {\bf 94}, 148101 (2005).

\bibitem{frolov06}
V. A. J. Frolov, Y. A. Chizmadzhev, F. S. Cohen, and J. Zimmerberg, Biophys. J. {\bf 91}, 189 (2006).

\bibitem{turner05}
M. S. Turner, P. Sens, and N. D. Socci, Phys. Rev. Lett. {\bf 95}, 168301 (2005).

\bibitem{lipowski92}
R. Lipowski, J. Phys. France II {\bf 2}, 1825 (1992).

\bibitem{julicher93}
F. J\"ulicher and R. Lipowski, Phys. Rev. Lett. {\bf 70}, 2964 (1993).

\bibitem{baumgart05}
T. Baumgart, S. Das, W. W. Webb, and J. T. Jenkins, Biophys. J. {\bf 89}, 1067 (2005).

\bibitem{siegel04}
D. P. Siegel and M. M. Kozlov, Biophys. J. {\bf 87}, 366 (2004).

\bibitem{allain04}
J.-M. Allain, C. Storm, A. Roux, M. Ben Amar, and J.-F. Joanny, Phys. Rev. Lett. {\bf 93}, 158104 (2004).

\bibitem{yanagisawa07}
M. Yanagisawa, M. Imai, T. Masui, S. Komura, and T. Otha, Biophys. J. {\bf 92}, 115 (2007).

\bibitem{reynwar07}
B. J. Reynwar, G. Illya, V. A. Harmandaris, M. M. M\"uller, K. Kremer, and M. Deserno, Nature {\bf 447}, 461 (2007).

\bibitem{baumgart03}
T. Baumgart, S. T. Hess, and W. W. Webb, Nature {\bf 425}, 821 (2003).

\bibitem{helfrich73}
P. B. Canham, J. Theor. Biol. {\bf 26}, 61 (1970); W. Helfrich, Z. Naturforsch. C {\bf 28}, 693 (1973).

\bibitem{variationalcalculus}
Hu Jian-Guo and Ou-Yang Zhong-Can, Phys. Rev. E {\bf 47}, 461 (1993); W.-M. Zheng and J. Liu, Phys. Rev. E {\bf 48}, 2856 (1993); F. J\"ulicher and U. Seifert, Phys. Rev. E {\bf 49}, 4728 (1994).

\bibitem{carmo76}
M. do Carmo, \emph{Differential Geometry of Curves and Surfaces}, (Prentice-Hall, Englewood Cliffs, NJ, 1976).

\bibitem{julicher96}
F. J\"ulicher and R. Lipowski, Phys. Rev. E {\bf 53}, 2670 (1996).

\bibitem{brakke92}
K. Brakke, Exper. Math. {\bf 1}, 141 (1992).

\bibitem{APLstuff}
The DOPC (1,2-di-oleoyl-sn-glycero-3-phosphocholine), sphingomyelin, cholesterol, Rhodamine-DOPE (1,2-dioleoyl-sn-glycero-3-phosphoethanolamine-N-(Lissamine Rhodamine B Sulfonyl)), and GM1 were obtained from Avanti Polar Lipids; the Alexa labeled choleratoxin from Molecular Probes.

\bibitem{mutz90}
M. Mutz and W. Helfrich, J. Phys. France {\bf 51}, 991 (1990).

\bibitem{pecreaux04}
J. P\'ecr\'eaux, H.-G. D\"obereiner, J. Prost, J.-F. Joanny, and P. Bassereau, Eur. Phys. J. E {\bf 13}, 277 (2004).

\bibitem{twoparameterfit}
A two-parameter fit without the continuity condition on~$\dot \psi$ at the domain boundary gives the same results within the experimental accuracy.

\bibitem{liu06}
J. Liu, M. Kaksonen, D. G. Drubin, and G. Oster, PNAS {\bf 103}, 10277 (2006).

\bibitem{muller05}
M. M. M\"uller, M. Deserno, and J. Guven, Europhys. Lett. {\bf 69}, 482 (2005).

\bibitem{scaffoldingproteins}
H. T. McMahon and J. L. Gallop, Nature {\bf 438}, 590 (2005); J. Zimmerberg and M. M. Kozlov, Nature Rev. Mol. Cell Biol. {\bf 7}, 9 (2006).

\end{thebibliography}

\end{document}